\documentclass[aps,preprint,prd,showpacs,nofootinbib]{revtex4}

\usepackage{amsmath}
\usepackage{graphicx}
\usepackage{dcolumn}
\usepackage{bm}
\usepackage{amssymb}
\usepackage{latexsym}
\usepackage{simplewick}
\usepackage{slashed,graphicx,color,amsmath,amssymb,simplewick}

\begin{document}

\title{Gravitational waves induced by massless vector fields with non-minimal coupling to gravity}

\author{Kaixi Feng\footnote{Email: fengkaixi10@mails.ucas.ac.cn}}

\affiliation{Institute of Theoretical Physics, Chinese Academy of Sciences, Beijing 100190, China}

\begin{abstract}
In this paper, we calculate the contribution of the late time mode of a massless vector field to the power spectrum of the primordial gravitational wave using retarded Green's propagator. We consider a non-trivial coupling between gravity and the vector field. We find that the correction is scale-invariant and of order $\frac{H^4}{M_P^4}$. The non-minimal coupling leads to a dependence of $\frac{H^2}{M^2}$, which can amplify the correlation function up to the level of $\frac{H^2}{M^2_P}$.

\end{abstract}
\maketitle

\section{Introduction}

General relativity as a combination of field theory and geometry \cite{vanHolten:2016avu}, has been fascinating us with its marvelous achievements and richness even hundred years after its birth. According to General relativity, gravity is the interaction between matter and space-time geometry. In this dynamical description of the space-time geometry, gravitation waves(GW) play an important role of storing and transporting energy and momentum. In this way, GW can be produced via the transformation of the space-time geometry itself and left an imprint in it. Therefore, through the study of GW, we might get rich information directly about new physics especially in the very early universe.

At the very early stage of our universe, it is supported that there has been through an exponential expansion which is arguably best described by inflation. Inflation models generally predict the generation of a stochastic background of primordial GW \cite{Grishchuk:1975ty, Starobinsky:1979ty, Rubakov:1982ty}. It might be able to shed light on the knowledge of our early universe if we are capable of getting information from the data of this GW background. The detection of primordial GW focuses on the observational signal of B-modes, a curl-like pattern that GW left in the Cosmic Microwave Background. Aiming at finding B-modes, there are a number of ongoing or forthcoming experiments, such as ACTPol \cite{Calabrese:2014gwa}, Polarbear \cite{Ade:2014afa}, Spider \cite{Crill:2008rd}, PRISM \cite{Andre:2013nfa}, LiteBIRD \cite{Matsumura:2013aja}. Beside the primordial GW from inflation, there are many other sources which can generate GW during the early stages of the universe, like cosmic deficits, phase transitions, and so on \cite{Damour:2000wa, Figueroa:2012kw, Kosowsky:1991ua, Kamionkowski:1993fg}.

In this paper, we are specifically interested in the GW sourced by vector fields, which is non-minimal coupled with curvature. See \cite{Jimenez:2013qsa, EspositoFarese:2009aj, deRham:2011by, Barrow:2012ay} for more about this kind of non-minimal coupling. For simplicity, we only consider the late time mode of a massless vector field as a first try. The case of massive vector fields will be considered in the coming future. This paper is organized as the following: In section II, we will derive the equation of motion of the vector field and the energy-momentum tensor for the vector-tensor theory we are considering; In section III, we will compute the GW induced by this vector field; Our conclusion is given in section IV.

We will use the following conventions: $sign(g_{\mu\nu})=(-,+,+,+)$,~$R_{\mu\nu}= R^\alpha_{\mu\alpha\nu}$,~$R^\mu_{~\nu\alpha\beta}=\Gamma^\mu_{\beta\nu,\alpha}-\Gamma^\mu_{\alpha\nu,\beta}+
    \Gamma^\mu_{\alpha\lambda}\Gamma^\lambda_{\beta\nu}-\Gamma^\mu_{\beta\lambda}\Gamma^\lambda_{\alpha\nu}$.

\section{A general action for a vector field with non-minimal coupling to gravity}
In this section, we will present the general action for a vector field with non-minimal coupling to gravity which gives rise to second-order equations of motion. we will derive the energy-momentum tensor and the equation of motion of the vector field.

The action is:
\begin{eqnarray}
\mathcal{S}&=&\int d^4x\sqrt{-g}\left(\frac{M_P^2}{2}R-\frac{1}{4}F^{\mu\nu}F_{\mu\nu}
   +\frac{1}{4M^2}L^{\alpha\beta\gamma\delta}F_{\alpha\beta}F_{\gamma\delta}\right)\nonumber\\
&=&\int d^4x\sqrt{-g}\left(\frac{M_P^2}{2}R-\frac{1}{4}F^{\mu\nu}F_{\mu\nu}+\frac{1}{2M^2}R_{\mu\nu\rho\sigma}\tilde{F}^{\mu\nu}\tilde{F}^{\rho\sigma}\right)
   \label{action},
\end{eqnarray}
where $L^{\alpha\beta\gamma\delta}=\frac{1}{2}\epsilon^{\alpha\beta\mu\nu}\epsilon^{\gamma\delta\rho\sigma}R_{\mu\nu\rho\sigma}$ is the dual of the Riemann tensor, and $\tilde{F}^{\mu\nu}=\frac{1}{2}\epsilon^{\mu\nu\alpha\beta}F_{\alpha\beta}$ the dual of $F_{\mu\nu}=\nabla_\mu A_\nu-\nabla_\nu A_\mu$, with $\epsilon^{\alpha\beta\mu\nu}=\epsilon^{[\alpha\beta\mu\nu]}$ the Levi-Civita tensor. $M^2$ denotes the non-minimal coupling between the vector field and gravity. See \cite{Jimenez:2013qsa} for more details of the construction of this action. As mention in \cite{Jimenez:2013qsa}, this action is constructed based on the divergent-free tensors in 4 dimensions, leaving the non-minimal coupling of the Horndeski interaction shown in the action eq.\eqref{action} the only interaction which involves the kinetic terms for a vector field and gives rise to second order equations of motion in 4 dimensions. Because of the divergent-free constraint and the antisymmetry property of $F_{\mu\nu}$, only $L^{\alpha\beta\gamma\delta}$ can be coupled to $F_{\mu\nu}F_{\alpha\beta}$, as shown in eq.\eqref{action}. While the other way of contraction of indices $L^{\alpha\beta\gamma\delta}F_{\alpha\gamma}F_{\beta\delta}$ is proportional to $L^{\alpha\beta\gamma\delta}F_{\alpha\beta}F_{\gamma\delta}$ because of the Bianchi identity:
\begin{eqnarray}
 &~&L^{\alpha\beta\gamma\delta}+L^{\alpha\gamma\delta\beta}+L^{\alpha\delta\beta\gamma}\nonumber\\
 &=&\frac{1}{2}[g^{\alpha\rho}g^{\beta\sigma}g^{\mu\gamma}g^{\nu\delta}-(\mu\leftrightarrow\sigma)-(\nu\leftrightarrow\sigma)]
    R_{\mu\nu\rho\sigma}\nonumber\\
 &=&\frac{1}{2}g^{\alpha\rho}g^{\beta\sigma}g^{\mu\gamma}g^{\nu\delta}(R_{\rho\sigma\mu\nu}+R_{\rho\mu\nu\sigma}+R_{\rho\nu\sigma\mu})=0~,
\end{eqnarray}
which leads to $L^{\alpha\beta\gamma\delta}F_{\alpha\gamma}F_{\beta\delta}= \frac{1}{2}L^{\alpha\gamma\beta\delta}F_{\alpha\gamma}F_{\beta\delta}$. So this proportionality can be reabsorbed in the coupling $M^2$.

Now we shall study the equation of motion of the vector field. Variation of the action eq.\eqref{action} by a small vector field is
\begin{eqnarray}
\delta\mathcal{S}&=&\int d^4x\sqrt{-g}\left(\nabla_\mu F^{\mu\nu}
   -\frac{1}{M^2}\nabla_\gamma(L^{\alpha\beta\gamma\nu}F_{\alpha\beta})\right)\delta A_\nu~,
\end{eqnarray}
from which we derive the equation of motion for the vector field:
\begin{eqnarray}
\nabla_\mu(F^{\mu\nu}-\frac{1}{M^2}L^{\mu\nu\alpha\beta}F_{\alpha\beta})
=(g^{\mu\alpha}g^{\nu\beta}-\frac{1}{M^2}L^{\mu\nu\alpha\beta})\nabla_\mu F_{\alpha\beta}=0~.
\end{eqnarray}
Within the FRW metric $ds^2=-dt^2+a^2(t)d\mathbf{x}^2$, the $0-$ and $i-$ components of the equation of motion of the vector field are
\begin{eqnarray}
 0&=&(-1+\frac{4H^2}{M^2})a^{-2}(\partial^i\partial_iA_0-\partial^i\dot{A}_i)\\
 0&=&(-1+\frac{4H^2}{M^2})a^{-2}[(\partial_0+H)(\dot{A}^i-\partial^iA_0)]
    +(1-\frac{4}{M^2}\frac{\ddot{a}}{a})a^{-4}[\partial^l(\partial_lA^i-\partial^iA_l)]\nonumber\\
 &~&+2(\frac{4\dot{H}}{M^2})a^{-2}H(\dot{A}^i-\partial^iA_0)~.
\end{eqnarray}
For photons and massive vector bosons in the transverse direction, we have $q_iA^i=0$, thus  $A_0=0$ from the $0-$ component of the field equation. While the $i-$ component gives:
\begin{eqnarray}
 0&=&(-1+\frac{4H^2}{M^2})[(\partial_0+H)\dot{A}^i]
    +(1-\frac{4}{M^2}\frac{\ddot{a}}{a})a^{-2}[\partial^l(\partial_lA^i-\partial^iA_l)]+\frac{8\dot{H}}{M^2}H\dot{A}^i~,
\end{eqnarray}
or in momentum space:
\begin{eqnarray}
\ddot{A}^i+C_1H\dot{A}^i+C_2\frac{q^2}{a^2}A^i=0~,
\end{eqnarray}
where
\begin{eqnarray}
C_1=\frac{1-\frac{4}{M^2}(H^2+2\dot{H})}{1-\frac{4H^2}{M^2}}~,\\
C_2=\frac{1-\frac{4}{M^2}(\dot{H}+H^2)}{1-\frac{4H^2}{M^2}}~.
\end{eqnarray}
The vector field can be written as
\begin{eqnarray}
 A_i(\mathbf{x},t)=\int d^3q\sum_\lambda\left[e^{i\mathbf{q}\cdot\mathbf{x}}e_i(\hat{q},\lambda)\alpha(\mathbf{q},\lambda)\mathcal{A}_q(t)
    +e^{-i\mathbf{q}\cdot\mathbf{x}}e_i^*(\hat{q},\lambda)\alpha^*(\mathbf{q},\lambda)\mathcal{A}^*_q(t)\right]~,
\end{eqnarray}
where $\lambda=1,2$ is the helicity index of a photon, $e_i(\hat{q},\lambda)$ is the polarization vector. The creation and annihilation operators satisfy usual commutation relations. The $\mathcal{A}_q(t)$ mode then satisfies the following equation:
\begin{eqnarray}
\mathcal{A}''_q-\mathcal{H}(1-C_1)\mathcal{A}'_q+C_2q^2\mathcal{A}_q=0~,
\end{eqnarray}
where $\mathcal{H}=\frac{a'}{a}=\dot{a}$, with $'$ being the derivative corresponding to the comoving time $\tau=\int\frac{dt}{a}$. For a de Sitter space, the solution for a massless vector field of general wavelength is exactly a plane wave
\begin{eqnarray}
\mathcal{A}_q(\tau)=\frac{1}{(2\pi)^{3/2}\sqrt{2q}}e^{-iq\tau}.
\end{eqnarray}

The energy-momentum tensor for the vector field can be derived from eq.\eqref{action}:
\begin{eqnarray}
 T_{\mu\nu}&=&\frac{2}{\sqrt{-g}}\frac{\delta\mathcal{S}}{\delta g^{\mu\nu}}\nonumber\\
 &=&-F_\mu^{~\rho}F_{\nu\rho}+\frac{1}{4}g_{\mu\nu}F^{\alpha\beta}F_{\alpha\beta}\nonumber\\
 &~&+\frac{1}{2M^2}[-R_{\alpha\beta\rho\sigma}\tilde{F}^{\alpha\beta}\tilde{F}^{\rho\sigma}g_{\mu\nu}
   +8R_{\mu\beta\rho\sigma}\tilde{F}_\nu^{~\beta}\tilde{F}^{\rho\sigma}
   +4\nabla_\rho\nabla_\sigma(\tilde{F}^{~\rho}_\nu\tilde{F}^{~\sigma}_\mu)]~,
\end{eqnarray}
with spacial components:
\begin{eqnarray}
 T_{ij}
 &=&-F_i^{~\rho}F_{j\rho}+\frac{1}{4}g_{ij}F^{\alpha\beta}F_{\alpha\beta}\nonumber\\
 &~&+\frac{1}{2M^2}[-R_{\alpha\beta\rho\sigma}\tilde{F}^{\alpha\beta}\tilde{F}^{\rho\sigma}g_{ij}
    +8R_{i\beta\rho\sigma}\tilde{F}_j^{~\beta}\tilde{F}^{\rho\sigma}
    +4\nabla_\rho\nabla_\sigma(\tilde{F}^{~\rho}_j\tilde{F}^{~\sigma}_i)]\label{T_ij}~.
\end{eqnarray}
In \eqref{appendix-B}, we show the derivation of the last term $\nabla_\rho\nabla_\sigma(\tilde{F}^{~\rho}_\nu\tilde{F}^{~\sigma}_\mu)$ in $T_{\mu\nu}$. Expanding the spacial components of the energy-momentum tensor eq.\eqref{T_ij} under the FRW metric gives
\begin{eqnarray}
T_{ij}=^{e}T_{ij}+~^{tt}T_{ij}+~^{xx}T_{ij}+~^{tx}T_{ij}~,
\end{eqnarray}
with the following explicit expression
\begin{eqnarray}
^{e}T_{ij}
&=&\dot{A}_i\dot{A}_j-a^{-2}(\partial_i A_k-\partial_kA_i)(\partial_j A^k-\partial^kA_j)
   -\frac{1}{2}\delta_{ij}\dot{A}_k\dot{A}^k+\frac{1}{2}a^{-2}\delta_{ij}(\partial_kA_l)(\partial^kA^l-\partial^lA^k)~,\nonumber\\
\end{eqnarray}
\begin{eqnarray}
~^{tt}T_{ij}&=&\frac{1}{2M^2}[4(\dot{H}+5H^2)\dot{A}_j\dot{A}_i-4(\dot{H}+4H^2)\dot{A}_k\dot{A}^k\delta_{ij}\nonumber\\
&~&-4H\delta_{ij}\dot{A}_k\ddot{A}^k-5a^{-2}\delta_{ij}(\partial_k\dot{A}_l)(\partial^k\dot{A}^l)
   +4a^{-2}\delta_{ij}(\partial_k\dot{A}_l)(\partial^l\dot{A}^k)\nonumber\\
&~&+a^{-2}(\partial_i\dot{A}_k)(\partial_j\dot{A}^k)+a^{-2}(\partial_k\dot{A}_j)(\partial^k\dot{A}_i)+2H\dot{A}_i\ddot{A}_j+2H\dot{A}_j\ddot{A}_i\nonumber\\
&~&+4a^{-2}(\partial_i\dot{A}^k-\partial^k\dot{A}_i)(\partial_j\dot{A}_k-\partial_k\dot{A}_j)]~,
\end{eqnarray}

\begin{eqnarray}
~^{xx}T_{ij}&=&\frac{1}{2M^2}[a^{-2}(28\dot{H}+10H^2)\delta_{ij}(\partial_kA_l)(\partial^kA^l-\partial^lA^k)\nonumber\\
&~&+4a^{-4}\delta_{ij}(\partial_m\partial_kA_l)(\partial^m\partial^kA^l-\partial^m\partial^lA^k)
   -4a^{-4}\delta_{ij}(\partial_k\partial^kA^n)(\partial^l\partial_lA_n)\nonumber\\
&~&+a^{-2}(-32\dot{H}-14H^2)(\partial_iA_k-\partial_kA_i)(\partial_jA^k-\partial^kA_j)\nonumber\\
&~&+4a^{-4}(\partial^n\partial_kA_j)(\partial_n\partial_iA^k)+4a^{-4}(\partial^n\partial_jA_k)(\partial_n\partial^kA_i)
   -8a^{-4}(\partial_i\partial_kA_e)(\partial_j\partial^kA^e)\nonumber\\
&~&-4a^{-4}(\partial_n\partial_kA_j)(\partial^n\partial^kA_i)
   +8a^{-4}(\partial_i\partial_jA_k)(\partial^l\partial_lA^k)+4a^{-4}(\partial^k\partial_kA_i)(\partial_l\partial^lA_j)\nonumber\\
&~&-4a^{-4}(\partial_i\partial_kA_j)(\partial^l\partial_lA^k)-4a^{-4}(\partial_j\partial_kA_i)(\partial^l\partial_lA^k)
   +4a^{-4}(\partial_i\partial_kA_e)(\partial_j\partial^eA^k)]~,
\end{eqnarray}

\begin{eqnarray}
~^{tx}T_{ij}&=&\frac{1}{2M^2}[18a^{-2}H\delta_{ij}(\partial_kA_l)(\partial^k\dot{A}^l-\partial^l\dot{A}^k)
   +4a^{-2}\delta_{ij}\ddot{A}^n(\partial^l\partial_lA_n)-4a^{-2}H\delta_{ij}\dot{A}_k(\partial^l\partial_lA^k)\nonumber\\
&~&-2a^{-2}\ddot{A}_i(\partial^k\partial_kA_j)-2a^{-2}\ddot{A}_j(\partial^k\partial_kA_i)\nonumber\\
&~&+4a^{-2}H(\partial_i\partial_jA_k)\dot{A}^k-2H\dot{A}_k(\partial_j\partial^kA_i)-2a^{-2}H(\partial_i\partial_kA_j)\dot{A}^k\nonumber\\
&~&+2a^{-2}H(\partial_k\partial^kA_i)\dot{A}_j+2a^{-2}H(\partial_k\partial^kA_j)\dot{A}_i\nonumber\\
&~&-4a^{-2}(\partial_i\partial_jA_k)\ddot{A}^k+2a^{-2}(\partial_i\partial_kA_j)\ddot{A}^k+2a^{-2}(\partial_j\partial_kA_i)\ddot{A}^k\nonumber\\
&~&-9a^{-2}H(\partial_iA^k-\partial^kA_i)(\partial_j\dot{A}_k-\partial_k\dot{A}_j)
   -9a^{-2}H(\partial_i\dot{A}^k-\partial^k\dot{A}_i)(\partial_jA_k-\partial_kA_j)]~.
\end{eqnarray}
Despite the simple form of the action eq.\eqref{action} and the FRW metric, the expansion of the energy-momentum tensor is not so simple. We will use these expressions of $T_{ij}$ and the plane wave solution of the equation of motion for the vector field to calculate the two-point correlation function for GW in the next section.

\section{correlation function}
In this section we will study the GW induced by the vector field non-minimally coupled to gravity as shown in the previous section. Since we are dealing only with the tensor modes here, the FRW metric is perturbed into
\begin{eqnarray}
 ds^2=-dt^2+(\delta_{ij}+h_{ij})dx^idx^j~.
\end{eqnarray}
The equation of motion for the tensor perturbation $h_{ij}$, sourced by vector field at second-order, can be expressed in conformal time as
\begin{eqnarray}
 h''_{ij}+2\frac{a'}{a}h'_{ij}-\partial^2h_{ij}=\frac{2}{M_P^2} T_{ij}^{TT}~,\label{h_ij}
\end{eqnarray}
where $T_{ij}^{TT}=\Pi_{ij}^{~~lm}T_{lm} $, and $T_{ij}$ is the energy-momentum tensor of matter, and the
TT-projection tensor $\Pi_{ij}^{~~lm}\equiv
P_i^mP_j^l-\frac{1}{2}P_{ij}P^{lm}$, while $P_{ij}$ is the
projection operator defined by
\begin{eqnarray}
 P_{ij}=\delta_{ij}-\frac{k_ik_j}{|\mathbf{k}|^2},
\end{eqnarray}
where $k_i/|\mathbf{k}|$ denotes the propagation direction
of a plane wave.
The solution of eq.\eqref{h_ij} takes the following form
\begin{eqnarray}
 h_{ij}(\mathbf{k},\tau)=\frac{2}{M_P^2}\int d\tau'G_k(\tau,\tau')T_{ij}^{TT}(\mathbf{k},\tau'),\label{h_ij-1}
\end{eqnarray}
and $G_k(\tau,\tau')$ is the retarded propagator solving the
homogeneous transformation of eq.\eqref{h_ij}. Within a de Sitter
background $a(\tau)=-(H\tau)^{-1}$, $G_k(\tau,\tau')$ reads
\begin{eqnarray}
 G_k(\tau,\tau')=\frac{1}{k^3\tau'^2}[(1+k^2\tau\tau')\sin k(\tau-\tau')+k(\tau'-\tau)\cos k(\tau-\tau')]\Theta(\tau-\tau')~.\label{greenfunction}
\end{eqnarray}
We will use eqs. (\ref{h_ij-1}, \ref{greenfunction}) in the
following process. The two-point correlation function of the tensor perturbation is
\begin{eqnarray}
&~&\langle h_{ij}(\mathbf{k},\tau)h_{ij}(\mathbf{k}',\tau)\rangle\nonumber\\
&=&\frac{4}{M_P^4}\int d\tau'G_k(\tau,\tau')\int d\tau''G_k(\tau,\tau'')\int d^3\mathbf{p}\int d^3\mathbf{p}'\langle\Pi_{ijlm}(k)T^{lm}(\mathbf{p},\tau')\Pi_{ijno}(k')T^{no}(\mathbf{p}',\tau'')\rangle~. \label{eq_hh}\nonumber\\
\end{eqnarray}
We choose some terms from $\langle~^{e}T^{lm}(\mathbf{p},\tau')~^{e}T^{no}(\mathbf{p}',\tau'')\rangle$ to show the detail procedure of the contraction of indices,
\begin{eqnarray*}
&~&\Pi_{ijlm}(k)\Pi_{ijno}(k')\langle\dot{A}^l\dot{A}^m[\dot{A}^n\dot{A}^o+a^{-2}(p'^n A_f-p'_fA^n)((k'-p')^o A^f-(k'-p')^fA^o)\\
&~&-\frac{1}{2}\delta^{no}\dot{A}_f\dot{A}^f-\frac{1}{2}a^{-2}\delta^{no}p'_fA_s((k'-p')^fA^s-(k'-p')^sA^f)]\rangle\\
&=&\Big([\Pi^{ln}(\hat{p})\Pi^{mo}(\widehat{k-p})+\Pi^{lo}(\hat{p})\Pi^{mn}(\widehat{k-p})]
    -\frac{1}{2}\delta^{no}[\Pi^l_f(\hat{p})\Pi^{mf}(\widehat{k-p})+\Pi^{lf}(\hat{p})\Pi^m_f(\widehat{k-p})]\Big)\\
&~&\times\mathcal{\dot{A}}_p(\tau')\mathcal{\dot{A}}^*_{p}(\tau'')\mathcal{\dot{A}}_{k-p}(\tau')\mathcal{\dot{A}}^*_{k-p}(\tau'')\\
&~&+a^{-2}[p^n(k-p)^o\Pi^l_f(\hat{p})\Pi^{mf}(\widehat{k-p})+(k-p)^np^o\Pi^{lf}(\hat{p})\Pi^m_f(\widehat{k-p})\\
&~&-p^n(k-p)^f\Pi^l_f(\hat{p})\Pi^{mo}(\widehat{k-p})-(k-p)^np^f\Pi^{lo}(\hat{p})\Pi^m_f(\widehat{k-p})\\
&~&-p_f(k-p)^o\Pi^{ln}(\hat{p})\Pi^{mf}(\widehat{k-p})-(k-p)_fp^o\Pi^{lf}(\hat{p})\Pi^{mn}(\widehat{k-p})\\
&~&+p_f(k-p)^f\Pi^{ln}(\hat{p})\Pi^{mo}(\widehat{k-p})+(k-p)_fp^f\Pi^{lo}(\hat{p})\Pi^{mn}(\widehat{k-p})\\
&~&-\delta^{no}p_f(k-p)^f\Pi^l_s(\hat{p})\Pi^{ms}(\widehat{k-p})+\frac{1}{2}p_f(k-p)^s\delta^{no}\Pi^l_s(\hat{p})\Pi^{mf}(\widehat{k-p})\\
&~&+\frac{1}{2}(k-p)_fp^s\delta^{no}\Pi^{lf}(\hat{p})\Pi^m_s(\widehat{k-p})]
   \mathcal{\dot{A}}_p(\tau')\mathcal{A}^*_{p}(\tau'')\mathcal{\dot{A}}_{k-p}(\tau')\mathcal{A}^*_{k-p}(\tau'')\\
&=&[1+\frac{(\mathbf{k}\cdot\mathbf{p})^2}{k^2p^2}+\frac{(\mathbf{k}\cdot\mathbf{q})^2}{k^2q^2}
   +\frac{(\mathbf{k}\cdot\mathbf{p})^2(\mathbf{k}\cdot\mathbf{q})^2}{k^4p^2q^2}]
   \mathcal{\dot{A}}_p(\tau')\mathcal{\dot{A}}^*_{p}(\tau'')\mathcal{\dot{A}}_{k-p}(\tau')\mathcal{\dot{A}}^*_{k-p}(\tau'')\\
&~&+4a^{-2}\frac{(\mathbf{k}\cdot\mathbf{p})(\mathbf{k}\cdot\mathbf{q})}{k^2}
   \mathcal{\dot{A}}_p(\tau')\mathcal{A}^*_{p}(\tau'')\mathcal{\dot{A}}_{k-p}(\tau')\mathcal{A}^*_{k-p}(\tau'')~,
\end{eqnarray*}
where
\begin{eqnarray}
\Pi_{ij}(\hat{p})\equiv\sum_{\lambda=1}^2e^*_i(\hat{p},\lambda)e_j(\hat{p},\lambda)=\delta_{ij}-\frac{p_ip_j}{|\mathbf{p}|^2}
\end{eqnarray}
for massless vector field. At the last step, and hereafter, we denote $\mathbf{q}\equiv\mathbf{k}-\mathbf{p}$. The contraction of indices is done by the mathematica package xAct \cite{xact}. The full contraction of $~^{e}T^{lm}~^{e}T^{no}$ is
\begin{eqnarray*}
&~&\Pi_{ijlm}(k)\Pi_{ijno}(k')\langle~^{e}T^{lm}(\mathbf{p},\tau')~^{e}T^{no}(\mathbf{p}',\tau'')\rangle\\
&=&[1+\frac{(\mathbf{k}\cdot\mathbf{p})^2}{k^2p^2}+\frac{(\mathbf{k}\cdot\mathbf{q})^2}{k^2q^2}
   +\frac{(\mathbf{k}\cdot\mathbf{p})^2(\mathbf{k}\cdot\mathbf{q})^2}{k^4p^2q^2}]
   \mathcal{\dot{A}}_p(\tau')\mathcal{\dot{A}}^*_{p}(\tau'')\mathcal{\dot{A}}_{k-p}(\tau')\mathcal{\dot{A}}^*_{k-p}(\tau'')\\
&~&+4a^{-2}\frac{(\mathbf{k}\cdot\mathbf{p})(\mathbf{k}\cdot\mathbf{q})}{k^2}
   \mathcal{\dot{A}}_p(\tau')\mathcal{A}^*_{p}(\tau'')\mathcal{\dot{A}}_{k-p}(\tau')\mathcal{A}^*_{k-p}(\tau'')\\
&~&+4a^{-2}\frac{(\mathbf{k}\cdot\mathbf{p})(\mathbf{k}\cdot\mathbf{q})}{k^2}
   \mathcal{A}_p(\tau')\mathcal{\dot{A}}^*_{p}(\tau'')\mathcal{A}_{k-p}(\tau')\mathcal{\dot{A}}^*_{k-p}(\tau'')\\
&~&+a^{-4}p^2q^2
   [1+\frac{(\mathbf{k}\cdot\mathbf{q})^2}{k^2q^2}+\frac{(\mathbf{k}\cdot\mathbf{p})^2}{k^2p^2}
   +\frac{(\mathbf{k}\cdot\mathbf{p})^2(\mathbf{k}\cdot\mathbf{q})^2}{k^4p^2q^2}]
   \mathcal{A}_p(\tau')\mathcal{A}^*_{p}(\tau'')\mathcal{A}_{k-p}(\tau')\mathcal{A}^*_{k-p}(\tau'')~.
\end{eqnarray*}
For simplicity, the rest of the calculation will consider the late time mode of the plane-wave solution, which is time independent.
Insert this into \eqref{eq_hh}, taking $\tau\to0$, integrating over time and momentums, we have
\begin{eqnarray}
&~&\langle h_{ij}(\mathbf{k})h_{ij}(\mathbf{k}')\rangle_{^eT^eT}\nonumber\\
&=&\frac{4\delta^3(\mathbf{k}+\mathbf{k}')}{M_P^4}\int d\tau'G_k(0,\tau')\int d\tau''G_k(0,\tau'')\int d^3\mathbf{p}\nonumber\\
&~&a^{-2}(\tau')a^{-2}(\tau'')p^2q^2
   [1+\frac{(\mathbf{k}\cdot\mathbf{q})^2}{k^2q^2}+\frac{(\mathbf{k}\cdot\mathbf{p})^2}{k^2p^2}
   +\frac{(\mathbf{k}\cdot\mathbf{p})^2(\mathbf{k}\cdot\mathbf{q})^2}{k^4p^2q^2}]
   \mathcal{A}_p(\tau')\mathcal{A}^*_{p}(\tau'')\mathcal{A}_{k-p}(\tau')\mathcal{A}^*_{k-p}(\tau'')\nonumber\\
&\sim&-\frac{8}{75}\frac{4\delta^3(\mathbf{k}+\mathbf{k}')}{(2\pi)^5}\frac{H^4}{M_P^4k^3}
~.
\end{eqnarray}
More details about the momentum integrals can be found in \cite{Feng:2015dxa}. The above result, and what follows in this paper, omitted divergent terms which should be renormalized away. We leave this renormalization problem in the future research.

Repeating the same procedure, the two-point correlation function coming from the $^eT^{xx}T$ and $^{xx}T^eT$ parts are:
\begin{eqnarray}
&~&\langle h_{ij}(\mathbf{k})h_{ij}(\mathbf{k}')\rangle_{^eT^{xx}T}\nonumber\\
&=&\frac{4\delta^3(\mathbf{k}+\mathbf{k}')}{M_P^4}\int d\tau'G_k(0,\tau')\int d\tau''G_k(0,\tau'')\int d^3\mathbf{p}\nonumber\\
&~&\frac{7a^{-2}(\tau')a^{-2}(\tau'')H^2}{M^2}p^2q^2[\frac{(\mathbf{k}\cdot\mathbf{p})^2(\mathbf{k}\cdot\mathbf{q})^2}{k^4p^2q^2}
  +\frac{(\mathbf{k}\cdot\mathbf{q})^2}{k^2q^2}+\frac{(\mathbf{k}\cdot\mathbf{p})^2}{k^2p^2}+1]
   \mathcal{A}_p(\tau')\mathcal{A}^*_{p}(\tau'')\mathcal{A}_{k-p}(\tau')\mathcal{A}^*_{k-p}(\tau'')\nonumber\\
&\sim&-\frac{8}{75}\frac{4\delta^3(\mathbf{k}+\mathbf{k}')}{(2\pi)^5}\frac{H^4}{M_P^4k^3}\frac{7H^2}{M^2}~,
\end{eqnarray}
and
\begin{eqnarray}
&~&\langle h_{ij}(\mathbf{k})h_{ij}(\mathbf{k}')\rangle_{^{xx}T^{e}T}\nonumber\\
&=&\frac{4\delta^3(\mathbf{k}+\mathbf{k}')}{M_P^4}\int d\tau'G_k(0,\tau')\int d\tau''G_k(0,\tau'')\int d^3\mathbf{p}\nonumber\\
&~&\{\frac{7a^{-2}(\tau')a^{-2}(\tau'')H^2}{M^2}p^2q^2[\frac{(\mathbf{k}\cdot\mathbf{p})^2(\mathbf{k}\cdot\mathbf{q})^2}{k^4p^2q^2}
  +\frac{(\mathbf{k}\cdot\mathbf{q})^2}{k^2q^2}+\frac{(\mathbf{k}\cdot\mathbf{p})^2}{k^2p^2}+1]\nonumber\\
&~&+a^{-4}(\tau')a^{-2}(\tau'')\frac{2}{M^2}[\frac{(\mathbf{k}\cdot\mathbf{p})^3(\mathbf{k}\cdot\mathbf{q})q^2}{k^4}
   -\frac{(\mathbf{k}\cdot\mathbf{p})(\mathbf{k}\cdot\mathbf{q})^3p^2}{k^4}+\frac{(\mathbf{k}\cdot\mathbf{p})^2(\mathbf{p}\cdot\mathbf{q})q^2}{k^2}
   -\frac{(\mathbf{k}\cdot\mathbf{q})^2(\mathbf{p}\cdot\mathbf{q})p^2}{k^2}]\}\nonumber\\
&~&\mathcal{A}_p(\tau')\mathcal{A}^*_{p}(\tau'')\mathcal{A}_{k-p}(\tau')\mathcal{A}^*_{k-p}(\tau'')\nonumber\\
&\sim&-\frac{8}{75}\frac{4\delta^3(\mathbf{k}+\mathbf{k}')}{(2\pi)^5}\frac{H^4}{M_P^4k^3}\frac{7H^2}{M^2}~.
\end{eqnarray}
Since we are considering the late time mode of the vector field, non-zero contribution comes from $^eT$ and $^{xx}T$ parts of the total energy-momentum tensor. Finally, summing up we have
\begin{eqnarray}
&~&\langle h_{ij}(\mathbf{k})h_{ij}(\mathbf{k}')\rangle_{massless,~late-time~mode}
\simeq-\frac{8}{75}\frac{4\delta^3(\mathbf{k}+\mathbf{k}')}{(2\pi)^5}\frac{H^4}{M_P^4k^3}\left(1+\frac{14H^2}{M^2}\right).
\end{eqnarray}
From the above result, depending on the realistic non-minimal coupling strength $M^2$, the non-minimal coupling effect might give rise to a sizable correction to the primordial GW of order $\frac{H^4}{M_P^4}\frac{14H^2}{M^2}$. For example, if the non-minimal coupling is strong enough to make $M^2\sim\frac{H^4}{M^2_P}$, the GW from this non-minimal coupling vector will be amplified to the level of $\frac{H^2}{M^2_P}$.

\section{Conclusion}
In this paper, we calculated the contribution of the late time mode of a massless vector field to the power spectrum of primordial gravitational wave using retarded Green's propagator. We considered a nontrivial coupling between gravity and the vector field with the form $\frac{1}{4M^2}L^{\alpha\beta\gamma\delta}F_{\alpha\beta}F_{\gamma\delta}$, which gives second-order equations of motion for both vector and tensor. We find that the correction is scale-invariant and of order $\frac{H^4}{M_P^4}$. Depending on the realistic non-minimal coupling strength $M^2$, the correlation function can be enhanced up to the level of $\frac{H^2}{M^2_P}$.

As a first try on calculating the non-minimal coupling between vector and the dual of the Riemann tensor, we find that the result is quite interesting despite the fact that we are considering the late time mode of a massless vector. We will study the massive vector which is non-minimal coupled to gravity as a further work.

\centerline{\textbf{Acknowledgments}}

K.F would like to thank Jiaming Zheng for helpful discussion.

\appendix

\section{Derivation of the energy-momentum tensor\label{appendix-B}}
In this appendix, we will use an example to show the derivation of the energy-momentum tensor from an action which is different from the Einstein gravity.
We will use the following action
\begin{eqnarray}
 S_{met}=\int_\mathcal{V}d^4x\sqrt{-g}f(G_2),
\end{eqnarray}
where $G_2\equiv R_{\mu\nu\rho\sigma}R^{\mu\nu\rho\sigma}$. Variation of the above action gives
\begin{eqnarray}
 \delta S_{met}&=&\int_\mathcal{V}d^4x[f(G_2)\delta\sqrt{-g}+\sqrt{-g}f'(G_2)\delta G_2]\nonumber\\
 &=&\int_\mathcal{V}d^4x[-\frac{1}{2}f(G_2)\sqrt{-g}g_{\alpha\beta}\delta g^{\alpha\beta}
    +\sqrt{-g}f'(G_2)(4R_{\mu_1\nu\rho\sigma}R_{\mu_2}^{~~\nu\rho\sigma}\delta g^{\mu_1\mu_2}+2R^{\mu\nu\rho\sigma}\delta R_{\mu\nu\rho\sigma})]\nonumber\\
 &=&\int_\mathcal{V}d^4x\sqrt{-g}\{[-\frac{1}{2}f(G_2)g_{\alpha\beta}+4f'(G_2)R_{\alpha\nu\rho\sigma}R_\beta^{~\nu\rho\sigma}]\delta g^{\alpha\beta}
    +f'(G_2)2R^{\mu\nu\rho\sigma}\delta R_{\mu\nu\rho\sigma}\}\nonumber\\
 &=&\int_\mathcal{V}d^4x\sqrt{-g}\{[-\frac{1}{2}f(G_2)g_{\alpha\beta}+4f'(G_2)R_{\alpha\nu\rho\sigma}R_\beta^{~\nu\rho\sigma}]\delta g^{\alpha\beta}
    +f'(G_2)2R_\mu^{\nu\rho\sigma}\delta R^\mu_{\nu\rho\sigma}\}.
\end{eqnarray}

While the variation $\delta R^\mu_{\nu\rho\sigma}$ is given by
\begin{eqnarray}
 \delta R^{\mu}_{\nu\rho\sigma}&=&\delta\Gamma^\mu_{\sigma\nu;\rho}-\delta\Gamma^\mu_{\rho\nu;\sigma}\nonumber\\
 &=&-\frac{1}{2}[g_{\sigma\alpha}\nabla_\rho\nabla_\nu(\delta g^{\alpha\mu})+g_{\nu\alpha}\nabla_\rho\nabla_\sigma(\delta g^{\alpha\mu})
    -g^{\mu\gamma}g_{\sigma\alpha}g_{\nu\beta}\nabla_\rho\nabla_\gamma(\delta g^{\alpha\beta})]\nonumber\\
 &~&+\frac{1}{2}[g_{\rho\alpha}\nabla_\sigma\nabla_\nu(\delta g^{\alpha\mu})+g_{\nu\alpha}\nabla_\sigma\nabla_\rho(\delta g^{\alpha\mu})
    -g^{\mu\gamma}g_{\rho\alpha}g_{\nu\beta}\nabla_\sigma\nabla_\gamma(\delta g^{\alpha\beta})].
\end{eqnarray}
Now the variation of the action becomes
\begin{eqnarray}
 \delta S_{met}&=&\int_\mathcal{V}d^4x\sqrt{-g}\{[-\frac{1}{2}f(G_2)g_{\alpha\beta}+4f'(G_2)R_{\alpha\nu\rho\sigma}R_\beta^{~\nu\rho\sigma}]\delta g^{\alpha\beta}\nonumber\\
 &~&+2f'(G_2)R_\mu^{\nu\rho\sigma}[g_{\rho\alpha}\nabla_\sigma\nabla_\nu(\delta g^{\alpha\mu})+g_{\nu\alpha}\nabla_\sigma\nabla_\rho(\delta g^{\alpha\mu})
    -g^{\mu\gamma}g_{\rho\alpha}g_{\nu\beta}\nabla_\sigma\nabla_\gamma(\delta g^{\alpha\beta})]\}.\label{variation}
\end{eqnarray}
The three terms in the second line of \eqref{variation} needs more calculation. Since the logic is exactly the same, we use the first term as an example
\begin{eqnarray}
 \int_\mathcal{V}d^4x\sqrt{-g}2f'(G_2)R_\mu^{\nu\rho\sigma}g_{\rho\alpha}\nabla_\sigma\nabla_\nu(\delta g^{\alpha\mu}).
\end{eqnarray}
Define $U^\sigma$ as
\begin{eqnarray}
 U^\sigma&\equiv&2f'(G_2)R_\mu^{\nu\rho\sigma}g_{\rho\alpha}\nabla_\nu(\delta g^{\alpha\mu})
    -2\delta g^{\alpha\mu}g_{\rho\alpha}\nabla_\nu(f'(G_2)R_\mu^{\nu\rho\sigma}),\\
 \nabla_\sigma U^\sigma&=&2\nabla_\nu[\nabla_\sigma(f'(G_2)R_\mu^{\nu\rho\sigma})g_{\rho\alpha}\delta g^{\alpha\mu}]
    -2\nabla_\nu\nabla_\sigma(f'(G_2)R_\mu^{\nu\rho\sigma})g_{\rho\alpha}\delta g^{\alpha\mu}
    +2f'(G_2)R_\mu^{\nu\rho\sigma}g_{\rho\alpha}\nabla_\sigma\nabla_\nu(\delta g^{\alpha\mu})\nonumber\\
 &~&-2\nabla_\sigma[\delta g^{\alpha\mu}g_{\rho\alpha}\nabla_\nu(f'(G_2)R_\mu^{\nu\rho\sigma})].
\end{eqnarray}
Then we have
\begin{eqnarray}
 &~&\int_\mathcal{V}d^4x\sqrt{-g}2f'(G_2)R_\mu^{\nu\rho\sigma}g_{\rho\alpha}\nabla_\sigma\nabla_\nu(\delta g^{\alpha\mu})\nonumber\\
 &=&\int_\mathcal{V}d^4x\sqrt{-g}\{\nabla_\sigma U^\sigma-2\nabla_\nu[\nabla_\sigma(f'(G_2)R_\mu^{\nu\rho\sigma})g_{\rho\alpha}\delta g^{\alpha\mu}]
    +2\nabla_\nu\nabla_\sigma(f'(G_2)R_\mu^{\nu\rho\sigma})g_{\rho\alpha}\delta g^{\alpha\mu}\nonumber\\
 &~&+2\nabla_\sigma[\delta g^{\alpha\mu}g_{\rho\alpha}\nabla_\nu(f'(G_2)R_\mu^{\nu\rho\sigma})]\}.
\end{eqnarray}
Repeating the same procedure, we have
\begin{eqnarray}
 \delta S_{met}
 &=&\int_\mathcal{V}d^4x\sqrt{-g}\{[-\frac{1}{2}f(G_2)g_{\alpha\beta}+4f'(G_2)R_{\alpha\nu\rho\sigma}R_\beta^{~\nu\rho\sigma}]\nonumber\\
 &~&+2\nabla_\nu\nabla_\sigma(f'(G_2)R_\beta^{\nu\rho\sigma})g_{\rho\alpha}+2\nabla_\rho\nabla_\sigma(f'(G_2)R_\beta^{\nu\rho\sigma})g_{\nu\alpha}
    -2\nabla_\gamma\nabla_\sigma(f'(G_2)R_\mu^{\nu\rho\sigma})g^{\mu\gamma}g_{\rho\alpha}g_{\nu\beta}\}\delta g^{\alpha\beta}\nonumber\\
 &~&+\int_\mathcal{V}d^4x\sqrt{-g}\{\nabla_\sigma U^\sigma-2\nabla_\nu[\nabla_\sigma(f'(G_2)R_\mu^{\nu\rho\sigma})g_{\rho\alpha}\delta g^{\alpha\mu}]
    +2\nabla_\sigma[\delta g^{\alpha\mu}g_{\rho\alpha}\nabla_\nu(f'(G_2)R_\mu^{\nu\rho\sigma})]\nonumber\\
 &~&+\nabla_\sigma V^\sigma-2\nabla_\rho[\nabla_\sigma(f'(G_2)R_\mu^{\nu\rho\sigma})g_{\nu\alpha}\delta g^{\alpha\mu}]
    +2\nabla_\sigma[\delta g^{\alpha\mu}g_{\nu\alpha}\nabla_\rho(f'(G_2)R_\mu^{\nu\rho\sigma})]\nonumber\\
 &~&-\nabla_\sigma W^\sigma+2\nabla_\gamma[\nabla_\sigma(f'(G_2)R_\mu^{\nu\rho\sigma})g^{\mu\gamma}g_{\rho\alpha}g_{\nu\beta}\delta g^{\alpha\beta}]
    -2\nabla_\sigma[\delta g^{\alpha\beta}g^{\mu\gamma}g_{\rho\alpha}g_{\nu\beta}\nabla_\gamma(f'(G_2)R_\mu^{\nu\rho\sigma})]\},
\end{eqnarray}
where
\begin{eqnarray}
 V^\sigma&\equiv&
    2f'(G_2)R_\mu^{\nu\rho\sigma}g_{\nu\alpha}\nabla_\rho(\delta g^{\alpha\mu})-2\delta g^{\alpha\mu}g_{\nu\alpha}\nabla_\rho(f'(G_2)R_\mu^{\nu\rho\sigma}),
\end{eqnarray}
\begin{eqnarray}
 W^\sigma&\equiv&2f'(G_2)R_\mu^{\nu\rho\sigma}g^{\mu\gamma}g_{\rho\alpha}g_{\nu\beta}\nabla_\gamma(\delta g^{\alpha\beta})
    -2\delta g^{\alpha\beta}g^{\mu\gamma}g_{\rho\alpha}g_{\nu\beta}\nabla_\gamma(f'(G_2)R_\mu^{\nu\rho\sigma}).
\end{eqnarray}
Using the Gauss-Stokes theorem and imposing the boundary condition $\delta g_{\alpha\beta}=0$, we have
\begin{eqnarray}
 \delta S_{met}
 &=&\int_\mathcal{V}d^4x\sqrt{-g}\{[-\frac{1}{2}f(G_2)g_{\alpha\beta}+4f'(G_2)R_{\alpha\nu\rho\sigma}R_\beta^{~\nu\rho\sigma}]\nonumber\\
 &~&+2\nabla_\nu\nabla_\sigma(f'(G_2)R_\beta^{\nu\rho\sigma})g_{\rho\alpha}+2\nabla_\rho\nabla_\sigma(f'(G_2)R_\beta^{\nu\rho\sigma})g_{\nu\alpha}
    -2\nabla_\gamma\nabla_\sigma(f'(G_2)R_\mu^{\nu\rho\sigma})g^{\mu\gamma}g_{\rho\alpha}g_{\nu\beta}\}\delta g^{\alpha\beta}\nonumber\\
 &~&+\oint_{\partial\mathcal{V}}d^3y\epsilon\sqrt{|h|}n_\sigma(U^\sigma+V^\sigma-W^\sigma)\label{variation-1},
\end{eqnarray}
where $h$ the determinant of the induced metric, $\epsilon$ is equal to +1 if $\partial\mathcal{V}$ is time-like and -1 if $\partial\mathcal{V}$ is
space-like. By choosing proper boundary terms, the last line in \eqref{variation-1} can be canceled. Since we are not interested in the boundary terms, we won't give further expressions for them. More details about the boundary terms have been given in \cite{Guarnizo:2010xr}.

Finally, we have
\begin{eqnarray}
 \frac{\delta S_{met}}{\delta g^{\alpha\beta}}
 &=&-\frac{1}{2}f(G_2)g_{\alpha\beta}+4f'(G_2)R_{\alpha\nu\rho\sigma}R_\beta^{~\nu\rho\sigma}
    +4\nabla_\nu\nabla_\sigma(f'(G_2)R_{\beta~\alpha}^{~\nu~\sigma}).
\end{eqnarray}

From this example, it is very easy to derive the last term in eq.\eqref{T_ij}.

\end{document}